
%
%
%
\input phyzzx
\tolerance=1000
\sequentialequations
\def\rl{\rightline}

\def\t1{{\tilde 1}}

\def\AEF{A.E. Faraggi}
\def\DVN{D. V. Nanopoulos}

\def\NPB#1#2#3{Nucl. Phys. B {\bf#1} (19#2) #3}
\def\PLB#1#2#3{Phys. Lett. B {\bf#1} (19#2) #3}

\def\IJMP#1#2#3{Int. J. Mod. Phys. A {\bf#1} (19#2) #3}

\REF\MMM{M. Leurer, Y. Nir and N. Seiberg, WIS-92/94/DEC--PH, and references
therein (to be published).}
\REF\GSW{M. Green, J. Schwarz and E. Witten,
Superstring Theory, 2 vols., Cambridge
University Press, 1987.}
\REF\EU{\AEF, \PLB{278}{92}{131}.}
\REF\TOP{\AEF, \PLB{274}{92}{47}.}
\REF\SLM{\AEF, \NPB{387}{92}{289}.}
\REF\DSW{M. Dine, N. Seiberg and E. Witten, \NPB{289}{87}{585}.}
\REF\KLN{S. Kalara, J. Lopez and D.V. Nanopoulos,
\PLB{245}{91}{421}; \NPB{353}{91}{650}.}
\REF\FFF{I. Antoniadis, C. Bachas, and C. Kounnas,
Nucl.Phys.{\bf B289}
(1987) 87; I. Antoniadis and C. Bachas,
Nucl.Phys.{\bf B298} (1988)
586; H. Kawai, D.C. Lewellen, and S.H.-H. Tye,
Phys.Rev.Lett.{\bf57} (1986)
1832; Phys.Rev.{\bf D34} (1986) 3794;
Nucl.Phys.{\bf B288} (1987) 1;
R. Bluhm, L. Dolan, and P. Goddard,
Nucl.Phys.{\bf B309} (1988) 330.}
\REF\REVAMP{I. Antoniadis, J. Ellis,
J. Hagelin, and \DVN, \PLB{231}{89}{65}.}
\REF\FNY{\AEF, D.V. Nanopoulos and K. Yuan, \NPB{335}{90}{437}.}
\REF\NRT{\AEF, WIS-92/48/NOV-PH.}
\REF\ADS{J.J. Atick, L.J. Dixon and A. Sen, \NPB{292}{87}{109};
         S. Cecotti, S. Ferrara and M. Villasante, \IJMP{2}{87}{1839}.}
\REF\GCU{A. Faraggi, WIS-92/17/NOV-PH (Phys. Lett. {\bf B}, in press).}
\REF\GLT{B. Grzadkowski, M. Lindner and S. Theisen,
\PLB{198}{87}{64}.}

\singlespace
\rl{WIS--93/3/JAN--PH}
\rl{\today}
\rl{T}
\normalspace
\smallskip
\titlestyle{\bf{Cabibbo Mixing in Superstring
Derived Standard--like Models}}
\author{Alon E. Faraggi
{\footnote*{e--mail address: fhalon@weizmann.bitnet}}
and Edi Halyo{\footnote\dag{e--mail address: jphalyo@weizmann.bitnet}}}
\smallskip
\centerline {Department of Physics, Weizmann Institute of Science}
\centerline {Rehovot 76100, Israel}
\titlestyle{ABSTRACT}

We examine the problem of generation mixing in realistic superstring
derived standard--like models, constructed in the free fermionic formulation.
We study the possible sources of family mixing in these models . In a
specific model we estimate the Cabibbo angle. We
argue that a Cabibbo angle of the correct
order of magnitude can be obtained in these models.

\pagenumber 1
\singlespace
\vskip 0.5cm
\endpage
\normalspace

\centerline{\bf 1. Introduction}

One of the fundamental problems in high energy physics is
the origin of the fermion masses and mixing hierarchy.
The standard model
uses thirteen free parameters to parameterize the observed spectrum.
Possible extensions to the standard model, like Grand Unified Theories
(GUTs) and supersymmetric GUTs, reduce the number of free
parameters and can explain inter--family relations between some of the
masses.
However, GUTs and SUSY GUTs can neither explain the hierarchy among the
generations nor the observed values for the family mixing.
Over the past few decades many attempts have been made
to understand the structure of fermion mass matrices in terms of
radiative corrections and additional horizontal symmetries that
constrain the allowed interactions [\MMM]. However, all these attempts
suffer from a large degree of arbitrariness.
Within the context of unified theories it is conceivable that
the free parameters in the fermion mass matrices are determined
by a fundamental theory at the Planck scale. Superstring
theories [\GSW] are the most developed Planck scale theories to date.
Therefore, it is important to examine whether realistic superstring
models can lead to a qualitative
understanding of the fermion mass matrices.

In Ref. [\EU,\TOP,\SLM] realistic superstring standard--like models
were constructed in the four dimensional free fermionic formulation,
with the following properties:
(1) Three and only three generations of chiral fermions.
(2) The gauge group is ${SU(3)_C}\times{SU(2)_L}\times{U(1)_{B-L}}\times
    {U(1)_{{T_3}_R}}\times U(1)^n\times{hidden}$.
    The weak hypercharge is uniquely given by
    $U(1)_Y=T_{3_R}+{1\over2}(B-L)$ and has the standard
    $SO(10)$ embedding. Therefore, it leads unambiguously to the prediction
    $\sin\theta_W^2={3\over8}$ at the unification scale.
(3) There are enough scalar doublets and singlets to break the symmetry in
    a realistic way and to generate a realistic fermion mass hierarchy
    [\TOP,\SLM].
(5) The models are free from gauge and gravitational anomalies apart from
    a single ``anomalous $U(1)_A$" symmetry that is broken by the
    Dine--Seiberg--Witten (DSW) mechanism [\DSW].
(6) The free fermionic standard--like models suggest an explanation for the
    fermion mass hierarchy. At the cubic level of the superpotential
    only the top quark gets a nonvanishing mass term. The mass terms for
    the lighter quarks and leptons are obtained from nonrenormalizable
    terms. $SO(10)$ singlet fields in these terms obtain nonvanishing
    VEVs by the application of the DSW mechanism. Thus, the order $N$
    nonrenormalizable terms, of the form $cffh(\Phi/M)^{N-3}$,
    become effective trilinear terms, where $f,h,\Phi$ denote fermions, scalar
    doublets and scalar singlets, respectively. $M$ is a Planck scale mass
    to be defined later. The effective Yukawa couplings are given by
    $\lambda=c(\langle \Phi \rangle/M)^{N-3}$
    where the calculable coefficients $c$ are of order one [\KLN].


In this paper we examine the problem  of generation mixing in the
realistic superstring derived
standard--like models. We show that the family mixing arises due to hidden
sector states that are obtained from specific sectors in the massless spectrum.
We contemplate two possible scenarios for generating the Cabibbo angle
in these models. One is due to condensates of a non--Abelian hidden gauge
group.
In the second scenario the hidden sector states obtain VEVs by the
application of the DSW mechanism. We demonstrate, in a specific model, that
the second scenario can produce a Cabibbo angle of the correct
order of magnitude, while the first scenario is marginal.

\bigskip
\centerline{\bf 2. The superstring standard--like models}

The superstring standard--like models are constructed in the four
dimensional free fermionic formulation [\FFF].
The models are generated by a basis of eight boundary condition vectors
for all world--sheet fermions.
The first five vectors in the basis
consist of the NAHE
set $\{{\bf 1},S,b_1,b_2,b_3\}$ [\REVAMP,\SLM].
The gauge group after the NAHE set is $SO(10)\times
E_8\times SO(6)^3$, with $N=1$ space--time supersymmetry,
and 48 spinorial $16$ of $SO(10)$.
The standard--like models are constructed by adding three additional
vectors to the NAHE set [\FNY,\EU,\TOP,\SLM].
Three additional vectors are needed to reduce the
number of generations to one generation from each sector $b_1$, $b_2$
and $b_3$. The three vectors that extend the NAHE set and the choice
of generalized GSO projection coefficients for our
model are given in table 1 [\EU].
The observable and hidden gauge groups after application
of the generalized GSO projections are
$SU(3)_C\times U(1)_C\times
 SU(2)_L\times U(1)_L\times U(1)^6${\footnote*{
$U(1)_C={3\over 2}U(1)_{B-L}$ and
$U(1)_L=2U(1)_{T_{3_R}}$.}}
and $SU(5)_H\times SU(3)_H\times U(1)^2$, respectively.
The weak hypercharge is given by
$U(1)_Y={1\over 3}U(1)_C + {1\over 2}U(1)_L$ and has the standard $SO(10)$
embedding. The orthogonal
combination is given by $U(1)_{Z^\prime}= U(1)_C - U(1)_L$.
The vectors $\alpha,\beta,\gamma$ break the $SO(6)_{j}$
horizontal symmetries to
$U(1)_{R_j}\times U(1)_{R_{j+3}}$ (j=1,2,3),
which correspond to the right--moving world--sheet
currents ${\bar\eta}^j_{1\over2}{{\bar\eta}^{j^*}}_{1\over2}$
($j=1,2,3$) and
${{\bar y}_3{\bar y}_6,{\bar y}_1{\bar\omega}_5,
{\bar\omega}_2{\bar\omega}_4}$, respectively.
For every right--moving $U(1)$ symmetry correspond
a left--moving global $U(1)$ symmetry. The first three
correspond to the charges of the supersymmetry generator
$\chi^{12}$, $\chi^{34}$ and $\chi^{56}$. The last three,
$U(1)_{\ell_{j+3}}$ $(j=1,2,3)$,
correspond to the complexified left--moving fermions
$y^3y^6$, $y^1\omega^5$ and $\omega^2\omega^4$.
Finally, the model contains six Ising model operators
that are obtained by pairing a left--moving
real fermion with a right--moving real fermion,
$\sigma^i_\pm=\{\omega^1{\bar\omega}^1,
y^2{\bar y}^2, \omega^3{\bar\omega}^3, y^4{\bar y}^4,
y^5{\bar y}^5, \omega^6{\bar\omega}^6\}_\pm$.

The full massless spectrum was presented in Ref. [\EU].
Here we list only the states that are relevant for the quark mass matrices.
The following massless states are produced by the sectors $b_{1,2,3}$,
$S+b_1+b_2+\alpha+\beta$, $O$ and their superpartners in the observable
sector:

(a) The $b_{1,2,3}$ sectors produce three $SO(10)$ chiral generations,
$G_\alpha=e_{L_\alpha}^c+u_{L_\alpha}^c+N_{L_\alpha}^c+d_{L_\alpha}^c+
Q_\alpha+L_\alpha$ $(\alpha=1,\cdots,3)$ where
$$\eqalignno{{e_L^c}&\equiv [(1,{3\over2});(1,1)];{\hskip .6cm}
{u_L^c}\equiv [({\bar 3},-{1\over2});(1,-1)];{\hskip .2cm}
Q\equiv [(3,{1\over2});(2,0)]{\hskip 2cm}
&(1a,b,c)\cr
{N_L^c}&\equiv [(1,{3\over2});(1,-1)];{\hskip .2cm}
{d_L^c}\equiv [({\bar 3},-{1\over2});(1,1)];{\hskip .6cm}
L\equiv [(1,-{3\over2});(2,0)]{\hskip 2cm}
&(1d,e,f)\cr}$$
of $SU(3)_C\times U(1)_C\times SU(2)_L\times U(1)_L$, with charges under the
six horizontal $U(1)$s,
$$\eqalignno{&({e_L^c}+{u_L^c})_{{1\over2},0,0,{1\over2},0,0}+
({d_L^c}+{N_L^c})_{{1\over2},0,0,{{1\over2}},0,0}+
(L)_{{1\over2},0,0,-{1\over2},0,0}+(Q)_{{1\over2},0,0,-{1\over2},0,0},
&(2a)\cr
&({e_L^c}+{u_L^c})_{0,{1\over2},0,0,{1\over2},0}+
({N_L^c}+{d_L^c})_{0,{1\over2},0,0,{1\over2},0}+
(L)_{0,{1\over2},0,0,-{1\over2},0}+
(Q)_{0,{1\over2},0,0,-{1\over2},0},
&(2b)\cr
&({e_L^c}+{u_L^c})_{0,0,{1\over2},0,0,{1\over2}}+
({N_L^c}+{d_L^c})_{0,0,{1\over2},0,0,{1\over2}}+
(L)_{0,0,{1\over2},0,0,-{1\over2}}+(Q)_{0,0,{1\over2},0,0,-{1\over2}}.
&(2c)\cr}$$
The vectors $b_1,b_2,b_3$ are the only vectors in the additive group
$\Xi$ which give rise to spinorial $16$ of $SO(10)$.

(b) The ${S+b_1+b_2+\alpha+\beta}$ sector gives
$$\eqalignno{h_{45}&\equiv{[(1,0);(2,1)]}_
{-{1\over2},-{1\over2},0,0,0,0} {\hskip .5cm}
D_{45}\equiv{[(3,-1);(1,0)]}_
{-{1\over2},-{1\over2},0,0,0,0}&(3a,b)\cr
\Phi_{45}&\equiv{[(1,0);(1,0)]}_
{-{1\over2},-{1\over2},-1,0,0,0}  {\hskip .5cm}
\Phi^{\pm}_1\equiv{[(1,0);(1,0)]}_
{-{1\over2},{1\over2},0,\pm1,0,0}&(3c,d)\cr
\Phi^{\pm}_2&\equiv{[(1,0);(1,0)]}_
{-{1\over2},{1\over2},0,0,\pm1,0} {\hskip .5cm}
\Phi^{\pm}_3\equiv{[(1,0);(1,0)]}_
{-{1\over2},{1\over2},0,0,0,\pm1}&(3e,f)\cr}$$
(and their conjugates ${\bar h}_{45}$, etc.).
The states are obtained by acting on the vacuum
with the fermionic oscillators
${\bar\psi}^{4,5},{\bar\psi}^{1,...,3},{\bar\eta}^3,{\bar y}^3\pm
i{\bar y}^6,{\bar y}^1\pm{i{\bar\omega}^5},
{\bar\omega}^2{\pm}i{\bar\omega}^4$,
respectively  (and their complex conjugates for ${\bar h}_{45}$, etc.).

(c) The Neveu--Schwarz $O$ sector gives, in addition to  the graviton,
dilaton, antisymmetric tensor and spin 1 gauge bosons,  scalar
electroweak doublets and singlets:
$$\eqalignno{{h_1}&\equiv{[(1,0);(2,-1)]}_{1,0,0,0,0,0}
{\hskip 2cm}\Phi_{23}\equiv{[(1,0);(1,0)]}_{0,1,-1,0,0,0}&(4a,b)\cr
{h_2}&\equiv{[(1,0);(2,-1)]}_{0,1,0,0,0,0}
{\hskip 2cm}\Phi_{13}\equiv{[(1,0);(1,0)]}_{1,0,-1,0,0,0}&(4c,d)\cr
{h_3}&\equiv{[(1,0);(2,-1)]}_{0,0,1,0,0,0}
{\hskip 2cm}\Phi_{12}\equiv{[(1,0);(1,0)]}_{1,-1,0,0,0,0}&(4e,f)\cr}$$
(and their conjugates ${\bar h}_1$, etc.).
Finally, the Neveu--Schwarz sector gives rise to three singlet
states that are neutral under all the U(1) symmetries.
$\xi_{1,2,3}:{\hskip .2cm}{\chi^{12}_{1\over2}{\bar\omega}^3_{1\over2}
{\bar\omega}^6_{1\over2}{\vert 0\rangle}_0},$
 ${\chi^{34}_{{1\over2}}{\bar y}_{1\over2}^5{\bar\omega}_{1\over2}^1
{\vert 0\rangle}_0},$
 $\chi^{56}_{1\over2}{\bar y}_{1\over2}^2{\bar y}_{1\over2}^4
{\vert 0\rangle}_0.$

The sectors $b_i+2\gamma+(I){\hskip .2cm} (i=1,..,3)$ give vector--like
representations that are
$SU(3)_C\times SU(2)_L\times {U(1)_L}\times {U(1)_C}$
singlets and transform as $5$, ${\bar 5}$ and $3$, ${\bar 3}$
under the
hidden $SU(5)$ and $SU(3)$ gauge groups, respectively (see table 2).
As will be shown below, the states from the sectors
$b_j+2\gamma$ produce the mixing between the chiral generations. We
would like to emphasize that the structure of the massless spectrum
exhibited in Eqs. (1--4), and in table 2, is common to a large number
of free fermionic standard--like models. All the standard--like models
contain three chiral generations from the sectors $b_j$, vector--like
representations from the sectors $b_j+2\gamma$, and Higgs doublets from
the Neveu--Schwarz sector and from the sector with
$\alpha+\beta$ plus some combination of $\{b_1,b_2,b_3\}$. Therefore,
the source of the family mixing is a general characteristic of these
models. It arises due to the basic set $\{{\bf 1},S,b_1,b_2,b_3\}$
and the use of the $Z_4$ twist to break the symmetry
from $SO(2n)$ to $SU(n)\times U(1)$.

In addition to the states above, the massless spectrum contains massless
states from sectors with some combination of
$\{b_1,b_2,b_3,\alpha,\beta\}$ and $\gamma+(I)$. These states are
model dependent and carry either fractional electric charge or
$U(1)_{Z^\prime}$ charge. As argued in Ref. [\NRT] the $U(1)_{Z^\prime}$
symmetry has to be broken at an intermediate energy scale that is
suppressed relative to the Planck scale. Therefore, the states from these
sectors do not play a significant role in the quark mass matrices and
we do not consider them in this paper.

The model contains six anomalous $U(1)$ symmetries:
Tr${U_1}=24$, Tr${U_2}=24$, Tr${U_3}=24$,
Tr${U_4}=-12$, Tr${U_5}=-12$, Tr${U_6}=-12$.
Of the six anomalous $U(1)$s,  five can be rotated by
an orthogonal transformation and one combination remains anomalous.
The six orthogonal combinations are given by [\EU],
$$\eqalignno{{U^\prime}_1&=U_1-U_2{\hskip .5cm},{\hskip .5cm}
{U^\prime}_2=U_1+U_2-2U_3,&(5a,b)\cr
{U^\prime}_3&=U_4-U_5{\hskip .5cm},{\hskip .5cm}
{U^\prime}_4=U_4+U_5-2U_6,&(5c,d)\cr
{U^\prime}_5&=U_1+U_2+U_3+2U_4+2U_5+2U_6,&(5e)\cr
U_A&=2U_1+2U_2+2U_3-U_4-U_5-U_6,&(5f)\cr}$$
with $Tr(Q_A)=180.$ The anomalous $U(1)$ symmetry
generates a large Fayet-Iliopoulos D--term by
the VEV of the dilaton field [\DSW]. Such a D--term would in general
break supersymmetry and destabilize the string  vacuum,
unless there is a direction
in the scalar potential $\phi={\sum_i}{\alpha_i\phi_i}$
 which is F--flat
and also D--flat with respect to the non--anomalous
gauge symmetries and in which
$\sum_i{Q_i^A{\vert\alpha_i\vert}^2}< 0$. If such a direction exists, it
will acquire a VEV, breaking the anomalous D--term, restoring supersymmetry
and stabilizing the vacuum [\ADS]. Since the fields
corresponding to such a flat
direction typically also carry charges for the nonanomalous D--terms,
a nontrivial set of constraints
on the possible choices of VEVs is imposed [\SLM].

\bigskip
\centerline{\bf 3. The superpotential}

We now turn to the superpotential of the model.
At the cubic level the following terms are obtained in the observable
sector [\EU],
$$\eqalignno{W_3&=\{(
{u_{L_1}^c}Q_1{\bar h}_1+{N_{L_1}^c}L_1{\bar h}_1+
{u_{L_2}^c}Q_2{\bar h}_2+{N_{L_2}^c}L_2{\bar h}_2+
{u_{L_3}^c}Q_3{\bar h}_3+{N_{L_3}^c}L_3{\bar h}_3)\cr
&\qquad
+{{h_1}{\bar h}_2{\bar\Phi}_{12}}
+{h_1}{\bar h}_3{\bar\Phi}_{13}
+{h_2}{\bar h}_3{\bar\Phi}_{23}
+{\bar h}_1{h_2}{\Phi_{12}}
+{\bar h}_1{h_3}{\Phi_{13}}
+{\bar h}_2{h_3}{\Phi_{23}}
+\Phi_{23}{\bar\Phi}_{13}{\Phi}_{12}\cr
&\qquad
+{\bar\Phi}_{23}{\Phi}_{13}{\bar\Phi}_{12}
+{\bar\Phi}_{12}({\bar\Phi}_1^+{\bar\Phi}_1^-
+{\bar\Phi}_2^+{\bar\Phi}_2^-
+{\bar\Phi}_3^+{\bar\Phi}_3^-)
+{\Phi_{12}}(\Phi_1^-\Phi_1^+
+\Phi_2^-\Phi_2^+
+\Phi_3^-\Phi_3^+)\cr
&\qquad
+{1\over2}\xi_3(\Phi_{45}{\bar\Phi}_{45}
+h_{45}{\bar h}_{45}
+D_{45}{\bar D}_{45}+\Phi_1^+{\bar\Phi}_1^++
\Phi_1^-{\bar\Phi}_1^-+\Phi_2^+{\bar\Phi}_2^++\Phi_2^-{\bar\Phi}_2^-
+\Phi_3^+{\bar\Phi}_3^+
\cr
&\qquad
+\Phi_3^-{\bar\Phi}_3^-)
+h_3{\bar h}_{45}\Phi_{45}+{\bar h}_3h_{45}{\bar\Phi}_{45}\}\quad&(6)\cr}$$
with a common normalization constant ${\sqrt 2}g$.

Nonrenormalizable contributions to the superpotential are obtained
by calculating corralators between vertex operators [\KLN],
$A_N\sim\langle V_1^fV_2^fV_3^b\cdot\cdot\cdot V_N^b\rangle,$
where $V_i^f$ $(V_i^b)$ are the fermionic (scalar)
components of the vertex operators.
In the analysis of nonrenormalizable terms we imposed the F--flatness
restriction $\langle{\bar\Phi}_{12},\Phi_{12},\xi_3\rangle\equiv0$
[\NRT].

At the quartic order there are no potential quark mass terms. At the
quintic order the following mass terms are obtained,
$$\eqalignno{&d_1Q_1h_{45}{\Phi}_1^+\xi_2
  {\hskip 2cm}d_2Q_2h_{45}{\bar\Phi}_2^-\xi_1 &(7a,b)\cr
             &u_1Q_1({\bar h}_{45}\Phi_{45}{\bar\Phi}_{13}+
  {\bar h}_2{\Phi}_i^+{\Phi}_i^-)&(7c)\cr
      &u_2Q_2({\bar h}_{45}\Phi_{45}{\bar\Phi}_{23}+
  {\bar h}_1{\bar\Phi}_i^+{\bar\Phi}_i^-)&(7d)\cr
      &(u_1Q_1h_1+u_2Q_2h_2)
  {{\partial W}\over{\partial\xi}_3}.&(7g)\cr}$$
At order $N=6$ we obtain mixing terms for $-{1\over3}$ charged quarks,
$$\eqalignno{
&d_3Q_2h_{45}\Phi_{45}V_3{\bar V_2},{\hskip 2cm}
d_2Q_3h_{45}\Phi_{45}V_2{\bar V_3},&(8a,b)\cr
&d_3Q_1h_{45}\Phi_{45}V_3{\bar V_1},{\hskip 2cm}
d_1Q_3h_{45}\Phi_{45}V_1{\bar V_3},&(8c,d)\cr}$$
At order $N=7$ we obtain in the down quark sector,
$$\eqalignno{
&d_2Q_1h_{45}\Phi_{45}(V_1{\bar V_2}+V_2{\bar V_1})\xi_i{\hskip .8cm}
d_1Q_2h_{45}\Phi_{45}(V_1{\bar V_2}+V_2{\bar V_1})\xi_i&(9a,b)\cr
&d_1Q_3h_{45}\Phi_{45}V_3{\bar V_1}\xi_2{\hskip 2.5cm}
d_3Q_1h_{45}\Phi_{45}V_1{\bar V_3}\xi_2&(9c,d)\cr
&d_2Q_3h_{45}\Phi_{45}V_3{\bar V_2}\xi_1{\hskip 2.5cm}
d_3Q_2h_{45}\Phi_{45}V_2{\bar V_3}\xi_1,&(9e,f) \cr}$$
where $\xi_i=\{\xi_1,\xi_2\}$.
In the up quark sector we obtain,
$$\eqalignno{
&u_1Q_2{\bar h}_1\Phi_{45}\{{\bar\Phi}_2^-(T_1{\bar T_2}+T_2{\bar T_1})+
                          {\bar\Phi}_1^+(V_1{\bar V_2}+V_2{\bar V_1}\}&(10a)\cr
&u_2Q_1{\bar h}_1\Phi_{45}\{{\bar\Phi}_1^-(T_1{\bar T_2}+T_2{\bar T_1})+
                          {\bar\Phi}_2^+(V_1{\bar V_2}+V_2{\bar V_1}\}&(10b)\cr
&u_1Q_2{\bar h}_2\Phi_{45}\{{\Phi}_2^+(T_1{\bar T_2}+T_2{\bar T_1})+
                            {\Phi}_1^-(V_1{\bar V_2}+V_2{\bar V_1}\}&(10c)\cr
&u_2Q_1{\bar h}_2\Phi_{45}\{{\Phi}_1^+(T_1{\bar T_2}+T_2{\bar T_1})+
                            {\Phi}_2^-(V_1{\bar V_2}+V_2{\bar V_1}\}&(10d)\cr
&u_3Q_1{\bar h}_1\Phi_{45}\{{\bar\Phi}_1^-T_1{\bar T_3}+
                            {\bar\Phi}_3^+V_3{\bar V_1}\}{\hskip .5cm}
 u_1Q_3{\bar h}_1\Phi_{45}\{{\bar\Phi}_3^-T_1{\bar T_3}+
                            {\bar\Phi}_1^+V_3{\bar V_1}\}&(10e)\cr
&u_3Q_1{\bar h}_2\Phi_{45}\{{\Phi}_3^+T_1{\bar T_3}+
                            {\Phi}_1^-V_3{\bar V_1}\}{\hskip .5cm}
 u_1Q_3{\bar h}_2\Phi_{45}\{{\Phi}_3^+T_1{\bar T_3}+
                            {\Phi}_1^-V_3{\bar V_1}\}&(10f)\cr
&u_3Q_2{\bar h}_1\Phi_{45}\{{\bar\Phi}_2^-T_2{\bar T_3}+
                            {\bar\Phi}_3^+V_3{\bar V_2}\}{\hskip .5cm}
 u_2Q_3{\bar h}_1\Phi_{45}\{{\bar\Phi}_3^-T_2{\bar T_3}+
                            {\bar\Phi}_2^+V_3{\bar V_2}\}&(10g)\cr
&u_3Q_2{\bar h}_2\Phi_{45}\{{\Phi}_2^+T_2{\bar T_3}+
                            {\Phi}_3^-V_3{\bar V_2}\}{\hskip .5cm}
 u_2Q_3{\bar h}_2\Phi_{45}\{{\Phi}_3^-T_2{\bar T_3}+
                            {\Phi}_2^-V_3{\bar V_2}\}&(10h)\cr}$$
At order $N=7$ we obtain generation mixing terms in the up and down quark
sectors. The states that induce the mixing come from the
sectors $b_j+2\gamma$. In the up quark sector, mixing is obtained
by $5$, ${\bar 5}$ and $3$, ${\bar 3}$ of the hidden $SU(5)$ and $SU(3)$
gauge groups, respectively. In the down quark sector, the mixing is only by
the $3$, ${\bar 3}$ of the hidden $SU(3)$ gauge groups.
At order $N=8$ we obtain mixing in the
down quark sector
by the $SU(5)$ states from the sectors $b_j+2\gamma$,
$$\eqalignno{
&d_3Q_1h_{45}\Phi_{45}\{{\Phi_1^+}{\bar\Phi}_3^-+
             {\Phi_3^+}{\bar\Phi}_1^-\}T_1{\bar T_3}{\hskip .5cm}
d_1Q_3h_{45}\Phi_{45}\{{\Phi_1^+}{\bar\Phi}_3^-+
             {\Phi_3^+}{\bar\Phi}_1^-\}T_3{\bar T_1}{\hskip 2mm}&(11a)\cr
&d_3Q_2h_{45}\Phi_{45}\{{\Phi_2^+}{\bar\Phi}_3^-+
             {\Phi_3^+}{\bar\Phi}_2^-\}T_2{\bar T_3}{\hskip .5cm}
d_2Q_3h_{45}\Phi_{45}\{{\Phi_2^+}{\bar\Phi}_3^-+
             {\Phi_3^+}{\bar\Phi}_2^-\}T_3{\bar T_2}{\hskip 2mm}&(11b) \cr}$$
The analysis of the nonrenormalizable terms up to order $N=8$
shows that family mixing terms are obtained for all generations. The
mixing arises due to the states from the sectors $b_j+2\gamma$. These
sectors, and their relation to the sectors $b_j$, is a general
characteristic of the realistic free fermionic models that use a $Z_4$ twist.
Therefore, the family mixing due to these states is a general
characteristic of these models.
In the next two sections we estimate the size of the off--diagonal terms, and
examine whether these models can account for the observed value of the
Cabibbo angle.

\bigskip
\centerline {\bf 4. Cabibbo mixing from hidden sector condensates }

The mixing terms in the previous section contained $5$, ${\bar 5}$ and
$3$, ${\bar 3}$ under the hidden $SU(5)$ and $SU(3)$ gauge groups,
respectively. These bilinears
may produce scalar condensates when $\alpha_h$ becomes large. More
generally, in any free fermionic standard--like model there is one
or several non--Abelian hidden gauge groups.
The largest
non--Abelian hidden gauge group that can be obtained in the
standard--like models is $SU(7)$.
Modifying the vector $\gamma$ of Ref. [\EU] by
$\gamma\{{\bar\phi}^3{\bar\phi}^4\}=1\rightarrow
\gamma\{{\bar\phi}^3{\bar\phi}^4\}=0$, enhances
the hidden gauge group from $SU(5)_H\times SU(3)_H\times U(1)^2$
to $SU(7)_H\times  U(1)^2$, for an appropriate choice of the generalized
GSO projection coefficients. The observable massless spectrum, Eqs. (1--4),
remains essentially the same. The states in the sectors $b_j+2\gamma$
form $7$ and ${\bar 7}$, and singlets, of $SU(7)_h$.
The anomalous $U(1)$s, and the anomaly free combinations are as in Eq. (5).
The cubic and quintic level terms are the same as in Eqs. (6--7).
The mixing terms are generated by $7$ and ${\bar 7}$ of the hidden
$SU(7)$ gauge group, and by the $SU(7)$ singlets, from the sectors
$b_j+2\gamma$. The mixing terms are similar to the terms in Eqs. (8--10).
Bilinear condensates
of states from the sectors $b_j+2\gamma$ that transform under the
hidden non--Abelian gauge groups may account for the generation mixing.
To estimate the possible magnitude
of mixing terms in the mass matrices we have to estimate the
condensates of the hidden $SU(7)$ gauge group.

The light Higgs representations are ${\bar h}_1$ or ${\bar h}_2$ and
$h_{45}$ [\NRT].
The mixing is dominantly in the down quark sector. An off--diagonal
term in the up--quark mass matrix does not contribute much to the
mixing because of the large diagonal terms.
The top quark mass term is obtained at the cubic level and the top
Yukawa coupling is of order one [\EU,\TOP].
Eq. (7) shows that both the bottom and
strange Yukawa couplings, as well as
the charm quark mass term,
can be obtained at the quintic order for an appropriate choice of singlet
VEVs [\NRT]. The VEVs of $\xi_1$ and $\xi_2$ are
undetermined and we use them to
fit the bottom and strange quark masses. We take $m_t\sim140$ $GeV$,
$\tan\beta={v_1/v_2}\sim1.5$, and therefore
$\lambda_b\sim0.01$,  which fixes
$\lambda_s\sim0.001$. The sector $b_3$ produces the lightest generation
states [\NRT]. Diagonal mass terms for the states from $b_3$ can only
be generated by VEVs that break $U(1)_{Z^\prime}$. We assume that
$U(1)_{Z^\prime}$ is broken at an intermediate scale that is suppressed
relative to the $SO(10)$ singlet VEVs. Otherwise, the cubic level
F--flat solution is violated by higher order nonrenormalizable
terms [\NRT]. Consequently, we take the diagonal
mass term for the lightest generation to be zero.
The mixing term between the two lightest
generations should be of the order
$O(10^{-4})$ to produce a Cabibbo angle of the correct order of magnitude.
Thus, we have to examine terms that mix $\{d_3;Q_3\}$ with
$\{d_2;Q_2\}$ or $\{d_1;Q_1\}$.

The scale $\Lambda_7$
at which the $SU(7)$ gauge coupling constant $\alpha_7$ gets strong
is given
by $$\alpha_7(\Lambda_7)={\alpha_7(M)\over{C(1-(b/2\pi)\alpha_7(M)
\ln(\Lambda_7/M)}} \approx 1\eqno(12)$$
where $b=(n_f/2)-21$. The bilinear hidden sector condensates produces
a suppression factor that is given by
$$({\Lambda_7\over{M}})^2=\exp({{64\pi}\over{b}})\eqno(13)$$
where
$\alpha_7(M)\approx 1/17$ from string gauge coupling unification [\GCU].
The suppression factor depends on the number of $7$ and ${\bar 7}$
that are massless below the Planck scale. In our model there are
eight pairs of $7$ and ${\bar 7}$, which gives
$({\Lambda_7/{M}})^2\sim10^{-6}$, and suppresses the mixing
below the observed values. Assuming that all $7$ and ${\bar 7}$
receive mass at the Planck scale gives a suppression factor
$({\Lambda_7/{M}})^2\sim5\times10^{-5}$, which is still too small.
The only way that bilinear hidden sector condensates may produce
sizable mixing is if the gauge coupling at the unification scale
turns out to be of the order $\alpha_U\sim{1\over8}-{1\over{10}}$.
Thus, we conclude that family mixing via bilinear condensates
in these models is marginal. It is possible if the hidden gauge group
is large, like $SU(7)$, and for a rather large gauge coupling at the
unification scale. If the hidden gauge group is $SU(5)$ or $SU(3)$,
$(\Lambda_h/M)$ will be smaller, and the mixing terms will be suppressed
even more.
In the next section we estimate the off--diagonal
terms in the case that some of the hidden sector states obtain nonvanishing
VEVs. In this scenario it is possible to obtain Cabibbo
angle of the correct order of magnitude, independent of
the specific hidden gauge group.

\bigskip
\centerline {\bf 5. Cabibbo mixing from hidden sector VEVs}

An alternative to the scenario discussed in the previous section is
that some of the hidden sector states, from the sectors $b_j+2\gamma$,
receive VEVs by the cancellation of the ``anomalous" $U(1)$ D--term
equation. We therefore have to find F and D flat solutions that
contain nonvanishing VEVs for the states from the sectors $b_j+2\gamma$.
An explicit solution that satisfies all the F and D flatness constraints is
given by the following set of nonvanishing VEVs,
$$\{V_2,{\bar V}_3,\Phi_{45},\Phi_{23},{\bar\Phi}_{23},\Phi_{13},
{\bar\Phi}_{13},\Phi_1^+,\Phi_2^\pm,{\bar\Phi}_1^\pm,
{\bar\Phi}_2^\pm\,\xi_1,\xi_2\},\eqno(14)$$
with
$$\eqalignno{&\vert{V_2}\vert^2=\vert{\bar V}_3\vert^2={1\over5}
\vert\Phi_{45}\vert^2=\vert{\bar\Phi}_1^-\vert^2={{g^2}\over{16\pi^2}}
{1\over{\sqrt{2\alpha^\prime}}},&(15a)\cr
&3\vert{\Phi_2^+}\vert^2=3\vert{{\bar\Phi}_2^+}\vert^2=
\vert{\Phi_2^-}\vert^2=\vert{{\bar\Phi}_2^-}\vert^2=
{1\over4}\vert{\Phi_1^+}\vert^2={1\over4}\vert{{\bar\Phi}_1^+}\vert^2=
\vert{\bar\Phi}_{13}\vert^2,&(15b)\cr
&\vert{\Phi_{23}}\vert^2=\vert{\bar\Phi}_{23}\vert^2=
{1\over3}\vert{\bar\Phi}_{13}\vert^2,&(15c)\cr
&\vert\Phi_{13}\vert^2=\vert{\bar\Phi}_{13}
\vert^2-{g^2\over{8\pi^2}}{1\over{\sqrt{2\alpha^\prime}}}.&(15d)\cr}$$
In this solution the VEVs of $\xi_1$, $\xi_2$ and ${\bar\Phi}_{13}$
are undetermined and remain free parameters. The choice of VEVs,
Eq. (12), does not affect the Higgs mass matrix up to order $N=8$.
Therefore, the light Higgs representations are still ${\bar h}_1$ or
${\bar h}_2$ and $h_{45}$ [\NRT].

The up and down quark mass matrices are diagonalized
by bi--unitary transformations,
$$\eqalignno{&U_LM_uU_R^\dagger=D_u\equiv{\rm diag}(m_u,m_c,m_t),&(16a)\cr
	     &D_LM_dD_R^\dagger=D_d\equiv{\rm diag}(m_d,m_s,m_b),&(16b)\cr}$$
with the mixing matrix given by,
$$V=U_LD_L^\dagger.\eqno(17)$$
For our F and D flat solution the down quark mass matrix takes the form,
$$M_d\sim\left(\matrix
{&\epsilon
&{{V_2{\bar V}_3\Phi_{45}}\over{M^4}} &0\cr
&{{V_2{\bar V}_3\Phi_{45}\xi_1}\over{M^3}}
&{{{\bar\Phi}_2^-\xi_1}\over{M^2}} &0 \cr
&0 &0
&{{\Phi_1^+\xi_2}\over{M^2}}\cr}\right)v_2\eqno(18)$$
where we have used [\KLN]
${1\over2}g\sqrt{2\alpha^\prime}=\sqrt{8\pi}/M_{Pl}$, to define
$M\equiv M_{Pl}/2\sqrt{8\pi}\approx 1.2\times 10^{18}GeV$.
Following Ref. [\KLN], we assume that the
coefficients of the nonrenormalizable terms are of order one. The bottom and
strange quark masses can be fitted by giving appropriate VEVs to $\xi_1$
and $\xi_2$. The 12 and 21 entries are then determined by our F and D
flat solution. Inserting the numerical values for the VEVs of $V_2$
${\bar V}_3$ and $\Phi_{45}$ from Eq. (15a), we obtain
$${{V_2{\bar V}_3\Phi_{45}}\over{M^3}}={{\sqrt{5}g^6}\over{64\pi^3}}
\approx 2-3\times 10^{-4}.\eqno(19)$$
Note that the down
quark mass matrix is not symmetric.
Only the 12 entry in the down quark mass matrix
has to be of the order $O(10^{-4})$
to obtain a Cabibbo angle of the correct order of magnitude.
We can use the remaining free parameter
to set ${{\langle{\bar\Phi}_2^-\rangle}/{M}}\sim0.001$, which imposes
$\xi_1\sim1$.
Inserting the numerical values to the mass matrix,
with $\epsilon<<10^{-4}$ and $g\sim0.8$,
and performing numerical singular value decomposition,
we obtain for the mixing matrix
$$\vert V \vert\sim \left(\matrix {0.98&0.2&0 \cr
                                        0.2&0.98&0 \cr
                                        0&0&1 \cr } \right) \eqno(20)$$
The running from the unification scale to the weak scale does not
affect the Cabibbo angle by much [\GLT].
Thus, we conclude that a Cabibbo angle of the correct order of
magnitude can be obtained in this scenario.

\bigskip
\centerline {\bf 6. Conclusions}

In this paper we discussed the problem of generation mixing
in superstring derived standard--like models. These models are
constructed in the free fermionic formulation. They correspond to
models that are compactified on $Z_2\times Z_2$ orbifold with
``standard embedding", and use a $Z_4$ twist to break the
symmetry from $SO(2n)$ to $SU(n)\times U(1)$.
We showed that the source of the mixing are the states
from the sectors $b_j+2\gamma$. We believe that the source of the
mixing is a general characteristic of these models, and is a
consequence of the basic set $\{{\bf 1},S,b_1,b_2,b_3\}$ and the
use of the $Z_4$ twist.
We examined two possible
scenarios for producing the family mixing. One is based on matter
condensates of a non--Abelian hidden gauge group. The other is
based on giving nonvanishing VEVs to states from the
sectors $b_j+2\gamma$, by the cancellation of the anomalous $U(1)$
D--term equation. We estimated the Cabibbo angle in the two
scenarios and showed that in the second scenario a Cabibbo angle
of the correct order of magnitude can be obtained in these models.
Mixing between the two heaviest generations can be obtained by finding
F and D flat solutions with a non vanishing VEV for $V_1$.
We will expand upon the phenomenology derived from these models in
future publications.

\bigskip
\centerline{\bf Acknowledgments}

This work is supported in part by a  Feinberg School Fellowship.
We would like to thank Yossi Nir for useful discussions.
Edi Halyo would like to express his gratitude to Isak and Gulen Halyo
for their support during this work.

\refout

\vfill
\eject

\input tables.tex

\magnification=1000
\tolerance=1200



{\hfill
{\begintable
\  \ \|\ ${\psi^\mu}$ \ \|\ $\{{\chi^{12};\chi^{34};\chi^{56}}\}$\ \|\
{${\bar\psi}^1$, ${\bar\psi}^2$, ${\bar\psi}^3$,
${\bar\psi}^4$, ${\bar\psi}^5$, ${\bar\eta}^1$,
${\bar\eta}^2$, ${\bar\eta}^3$} \ \|\
{${\bar\phi}^1$, ${\bar\phi}^2$, ${\bar\phi}^3$, ${\bar\phi}^4$,
${\bar\phi}^5$, ${\bar\phi}^6$, ${\bar\phi}^7$, ${\bar\phi}^8$} \crthick
$\alpha$
\|\ 0 \|
$\{0,~0,~0\}$ \|
1, ~~1, ~~1, ~~0, ~~0, ~~0 ,~~0, ~~0 \|
1, ~~1, ~~1, ~~1, ~~0, ~~0, ~~0, ~~0 \nr
$\beta$
\|\ 0 \| $\{0,~0,~0\}$ \|
1, ~~1, ~~1, ~~0, ~~0, ~~0, ~~0, ~~0 \|
1, ~~1, ~~1, ~~1, ~~0, ~~0, ~~0, ~~0 \nr
$\gamma$
\|\ 0 \|
$\{0,~0,~0\}$ \|
{}~~$1\over2$, ~~$1\over2$, ~~$1\over2$, ~~$1\over2$,
{}~~$1\over2$, ~~$1\over2$, ~~$1\over2$, ~~$1\over2$ \| $1\over2$, ~~0, ~~1,
{}~~1,
{}~~$1\over2$,
{}~~$1\over2$, ~~$1\over2$, ~~0 \endtable}
\hfill}
\smallskip
{\hfill
{\begintable
\  \ \|\
${y^3y^6}$,  ${y^4{\bar y}^4}$, ${y^5{\bar y}^5}$,
${{\bar y}^3{\bar y}^6}$
\ \|\ ${y^1\omega^6}$,  ${y^2{\bar y}^2}$,
${\omega^5{\bar\omega}^5}$,
${{\bar y}^1{\bar\omega}^6}$
\ \|\ ${\omega^1{\omega}^3}$,  ${\omega^2{\bar\omega}^2}$,
${\omega^4{\bar\omega}^4}$,  ${{\bar\omega}^1{\bar\omega}^3}$  \crthick
$\alpha$ \|
1, ~~~0, ~~~~0, ~~~~0 \|
0, ~~~0, ~~~~1, ~~~~1 \|
0, ~~~0, ~~~~1, ~~~~1 \nr
$\beta$ \|
0, ~~~0, ~~~~1, ~~~~1 \|
1, ~~~0, ~~~~0, ~~~~0 \|
0, ~~~1, ~~~~0, ~~~~1 \nr
$\gamma$ \|
0, ~~~1, ~~~~0, ~~~~1 \|\
0, ~~~1, ~~~~0, ~~~~1 \|
1, ~~~0, ~~~~0, ~~~~0  \endtable}
\hfill}
\smallskip
\parindent=0pt
\hangindent=39pt\hangafter=1
\baselineskip=18pt

{{\it Table 1.} A three generations ${SU(3)\times SU(2)\times U(1)^2}$
model. The choice of generalized GSO coefficients is:
${c\left(\matrix{b_j\cr
                                    \alpha,\beta,\gamma\cr}\right)=
-c\left(\matrix{\alpha\cr
                                    1\cr}\right)=
c\left(\matrix{\alpha\cr
                                    \beta\cr}\right)=
-c\left(\matrix{\beta\cr
                                    1\cr}\right)=
c\left(\matrix{\gamma\cr
                                    1,\alpha\cr}\right)=
-c\left(\matrix{\gamma\cr
                                    \beta\cr}\right)=
-1}$ (j=1,2,3),
with the others specified by modular invariance and space--time
supersymmetry.
Trilevel Yukawa couplings are obtained only for
${+{2\over3}}$ charged quarks.
The $16$ right--moving
internal fermionic states
$\{{\bar\psi}^{1,\cdots,5},{\bar\eta}^1,
{\bar\eta}^2,{\bar\eta}^3,{\bar\phi}^{1,\cdots,8}\}$,
correspond to the $16$ dimensional compactified  torus of the ten dimensional
heterotic string.
The 12 left--moving and 12 right--moving real internal fermionic states
correspond to the six left
and six right compactified dimensions in the bosonic language.
$\psi^\mu$ are the two space--time
external fermions in the light--cone gauge and
$\chi^{12}$, $\chi^{34}$, $\chi^{56}$
correspond to the spin connection in the bosonic constructions.}
\vskip 2.5cm

\vfill
\eject

\input tables.tex
\magnification=1000
\baselineskip=18pt
\hbox
{\hfill
{\begintable
\ F \ \|\ SEC \ \|\ $SU(3)_C$ $\times$ $SU(2)_L$ \ \|\ $Q_C$ & $Q_L$ & $Q_1$ &
$Q_2$
 & $Q_3$ & $Q_4$  & $Q_5$ & $Q_6$
 \ \|\ $SU(5)$ $\times$ $SU(3)$ \ \|\ $Q_7$ &
$Q_8$  \crthick
$V_1$ \|\ ${b_1+2\gamma}+(I)$ \|(1,1)\|~0 & ~~0 & ~~0 & ~~${1\over 2}$ &
 ~~$1\over 2$ & ~~$1\over2$
 & ~~0 & ~~0 \|(1,3)\| $-{1\over 2}$ &
{}~~$5\over 2$   \nr
${\bar V}_1$ \|\                \|(1,1)\| ~~0 & ~~0 & ~~0 & ~~${1\over 2}$ &
 ~~$1\over 2$ & ~~$1\over2$
  & ~~0 & ~~0
  \|(1,$\bar 3$)\| ~~${1\over 2}$ &
$-{5\over 2}$  \nr
$T_1$ \|\                \|(1,1)\| ~~0 & ~~0 & ~~0 & ~~${1\over 2}$ &
 ~~$1\over 2$ & $-{1\over2}$
  & ~~0 & ~~0
  \|(5,1)\| $-{1\over 2}$ &
$-{3\over 2}$  \nr
${\bar T}_1$ \|\                \|(1,1)\| ~~0 & ~~0 & ~~0 & ~~${1\over 2}$ &
 ~~$1\over 2$ & $-{1\over2}$
   & ~~0 & ~~0
  \|($\bar 5$,1)\| ~~$1\over2$ &
{}~~${3\over 2}$  \cr
$V_{2}$ \|\ ${b_2+2\gamma}+(I)$ \|(1,1)\| ~~0 & ~~0 &
 ~~${1\over 2}$ & ~~0 &
 ~~$1\over 2$ & ~~0 &  ~~$1\over2$ & ~~0
 \|(1,3)\| $-{1\over 2}$ &
 ~~$5\over 2$  \nr
${\bar V}_{2}$ \|\                \|(1,1)\| ~~0 & ~~0 & ~~${1\over 2}$ & ~~0 &
 ~~$1\over 2$ & ~~0 & ~~$1\over2$ & ~~0
  \|(1,$\bar 3$)\| ~~${1\over 2}$ &
 $-{5\over 2}$  \nr
$T_{2}$ \|\                \|(1,1)\| ~~0 & ~~0 & ~~${1\over 2}$ & ~~0 &
 ~~$1\over 2$ & ~~0 & $-{1\over2}$ & ~~0
  \|(5,1)\| $-{1\over 2}$ &
 $-{3\over 2}$ \nr
${\bar T}_{2}$ \|\                \|(1,1)\| ~~0 & ~~0 & ~~${1\over 2}$ & ~~0 &
 ~~$1\over 2$ & ~~0 & $-{1\over2}$ & ~~0
  \|($\bar 5$,1)\| ~~$1\over 2$ &
 ~~${3\over 2}$  \cr
$V_{3}$ \|\ ${b_3+2\gamma}+(I)$ \|(1,1)\| ~~0 & ~~0 & ~~${1\over 2}$ &
 ~~${1\over 2}$ & ~~0 & ~~0 & ~~0 & ~~${1\over2}$
       \|(1,3)\| $-{1\over 2}$ &
 ~~$5\over 2$  \nr
${\bar V}_{3}$ \|\                \|(1,1)\| ~~0 & ~~0 & ~~${1\over 2}$ &
 ~~${1\over 2}$ & ~~0 & ~~0 & ~~0 & ~~$1\over2$
 \|(1,$\bar 3$)\| ~~${1\over 2}$
 & $-{5\over 2}$  \nr
$T_{3}$ \|\                \|(1,1)\| ~~0 & ~~0 & ~~${1\over 2}$ &
 ~~${1\over 2}$ & ~~0 & ~~0 & ~~0 & $-{1\over2}$
       \|(5,1)\|
 $-{1\over 2}$ & $-{3\over 2}$  \nr
${\bar T}_{3}$ \|\                \|(1,1)\| ~~0 & ~~0 & ~~${1\over2}$ &
 ~~${1\over 2}$ & ~~0 & ~~0 & ~~0 & $-{1\over2}$
   \|($\bar 5$,1)\| ~~$1\over 2$
 & ~~${3\over 2}$
 \endtable}
\hfill}
\bigskip
\parindent=0pt
\hangindent=39pt\hangafter=1
{\it Table 2.} Massless states from the sectors $b_j+2\gamma$,
and their quantum numbers.

\vfill
\eject

\end
\bye